\documentclass[12pt]{article}
\usepackage{graphicx}
\begin{document}

%----------------------------------------------------------------------

\begin{center}
{\Large \bf Internal Space-time Symmetries of Particles \\[0.5ex]
  derivable from \\[1ex]
   Periodic Systems in Optics}

\vspace{3ex}

Y. S. Kim \footnote{electronic address: yskim@physics.umd.edu}\\
Department of Physics, University of Maryland,\\
College Park, Maryland 20742, U.S.A.

\end{center}

\vspace{3ex}
\begin{abstract}

While modern optics is largely a physics of harmonic oscillators
and two-by-two matrices, it is possible to learn about some hidden
properties of the two-by-two matrix from optical systems. Since
two-by-two matrices can be divided into three conjugate classes
depending on their traces, optical systems force us to establish
continuity from one class to another.  It is noted that those
three classes are equivalent to three different branches of
Wigner's little groups dictating the internal space-time
symmetries massive, massless, and imaginary-mass particles.
It is shown that the periodic systems in optics can also be
described by have the same class-based matrix algebra.  The
optical system allow us to make continuous, but not analytic,
transitions from massiv to massless, and massless to
imaginary-mass cases.
\end{abstract}
% PACS: 02.20.Qs, 11.30.-j, 78.67.Qt

\section{Introduction}\label{intro}
Two-by-two matrices with real elements have three independent
parameters if their determinants are constrained to be one.  They
constitute building blocks for many branches of physics, including
beam-transfer matrices in optics~\cite{theo79} and Wigner's little
groups for internal space-time symmetry of
particles~\cite{wig39,knp86,kiwi90jm}.
\par
For the two-by-two matrix, we are accustomed to solve a quadratic
equation to get the eigenvalues and construct a rotation matrix
to get the eigenvalues.  This procedure does not always lead to
correct answers, because squeeze matrices should also be
considered~\cite{hkn99ajp}.  We are quite familiar with rotations,
but the concept of squeeze started getting our attention only
after squeezed states of light appeared in the physics
literature~\cite{yuen76}.
\par
In particle physics, Lorentz boosts are squeeze transformations,
and this aspect was addressed by Paul A. M. Dirac in his 1949
and 1963 papers~\cite{dir49,dir63}.  It is possible to apply this
concept to high-speed hadrons which are bound state of the quarks
which are thought to be more fundamental particles~\cite{knp86}.
Thus, high-energy hadronic physics and modern optics share the
same mathematical base, and it is profitable to trade physics
between these two branches of physics using the common mathematics
language.
\par
In this report, we note first that optical beam transfer matrix,
often called the $ABCD$ matrix, has three independent parameters
and its determinant is one.  We study then how this matrix can be
decomposed into one-parameter matrices.   The Bargmann decomposition
and Iwasawa decomposition are already familiar to us, and are used
often in the literature~\cite{barg47,gk01}.
\par
We show that, in addition, there is a decomposition based on the
concept of conjugate classes~\cite{hamer62}.  There are three
conjugate classes.  The first class consists of those matrices
with their traces smaller than two, the second class consists of
those with the traces equal to two, and the third consisting of
those matrices with their traces greater than two.  It is remarkable
that this purely mathematical theorem corresponds to Wigner's
construction of his little groups. which dictate the internal
space-time symmetries of massive, massless, and imaginary-mass
particles respectively~\cite{wig39}.

\par
In Sec.~\ref{decom}, we introduce the Wigner decomposition based
on the conjugate classes of the two-by-two matrices.  It is then
shown that the Wigner decomposition can be translated into
the Bargmann and Iwasawa decompositions.
In Sec.~\ref{optsys}, we discuss how we can formulate those three
decompositions while studying optical multilayer systems.  It is
shown that the Wigner decomposition is needed for repeated
application of the $ABCD$ matrix for periodic system.
 In Sec.~\ref{little}, we study how those three decompositions can
serve useful purposes in studying Wigner's little groups and thus
the internal space-time symmetries of elementary particles.

\section{Decompositions of the ABCD Matrix}\label{decom}
The two-by-two matrix with real elements and unit determinant can
be written as
\begin{equation}\label{abcd00}
\pmatrix{A & B \cr C & D} ,
\end{equation}
with $AD - BC = 1$ has three independent elements.  This matrix is
commonly used as the beam transfer matrices in optics~\cite{theo79}.
The complete set of these matrices is like the group $SU(1,1)$
which serves as the fundamental language for squeezed states of
light~\cite{yuen76,knp91}
\par
All the matrices in this set can be divided into three classes
depending on their traces~\cite{hamer62}. Without changing its
trace, we can bring every matrix to an equi-diagonal form by a
rotation~\cite{bk09}.

We shall use the notation $[ABCA]$ for the equi-diagonal $ABCD$
matrix.  This equi-diagonal matrix now has two-independent
parameters.  If its trace is smaller than two, the matrix can
be written as
\begin{equation}\label{wigd11}
[ABCA] = \pmatrix{\cos\phi & -e^{\eta}\sin\phi \cr
       e^{-\eta}\sin\phi & \cos\phi}.
\end{equation}
If the trace is greater than two, it can take the form
\begin{equation}\label{wigd22}
[ABCA] = \pmatrix{\cosh\chi & -e^{\eta}\sinh\chi \cr
      -e^{-\eta}\sinh\chi & \cosh\chi}.
\end{equation}
If the trace is equal to two, it can be brought to the form
\begin{equation}\label{wigd33}
[ABCA] = \pmatrix{1 & -\gamma e^{\eta} \cr 0 & 1}.
\end{equation} .
We choose to use the notation $W(\tau)$ collectively
for the following three matrices.
\begin{equation}\label{wigd55}
\pmatrix{\cos\phi & -\sin\phi \cr \sin\phi & \cos\phi}, \quad
\pmatrix{\cosh\chi & -\sinh\chi \cr -\sinh\chi & \cosh\chi}, \quad
\pmatrix{1 & -\gamma \cr 0 & 1} .
\end{equation}
The parameter $\tau$ could be $\theta, \chi,$ or $\gamma$. If we
define the matrix $B(\eta)$ as
\begin{equation}\label{wigd66}
B(\eta) = \pmatrix{e^{\eta/2} & 0 \cr 0 & e^{-\eta/2}} .
\end{equation}
\par
Then every equi-diagonal $ABCA$ matrix can be written as
\begin{equation}\label{wigd77}
 [ABCA] = B(\eta) W(\tau) B(-\eta) .
\end{equation}
Indeed, the matrix $W(\tau)$ constitutes a set of three matrices
which play an important role in the theory of two-by-two matrices.
\par
It is important to note that these three matrices constitute the
basic elements for Wigner's little group which dictates the internal
space-time symmetries of elementary particles~\cite{wig39,knp86}.
We thus choose to call $W(\tau)$ the ``Wigner matrix''~\cite{bk10jmo}
and call the expression of Eq.(\ref{wigd77}) ``Wigner decomposition."
The Wigner decomposition leads to
\begin{equation}
 [ABCA]^N = B(\eta)[W(\tau)]^N B(-\eta) = B(\eta) W(N\tau) B(-\eta) ,
\end{equation}
convenient in dealing with repeated applications of the $ABCA$ matrix
in periodic systems.

It is known that the $ABCA$ matrix can be written as the product of
three one-parameter matrices
\begin{equation}\label{bargd22}
\pmatrix{\cos(\theta/2) & \sin(\theta/2) \cr-\sin(\theta/2) &
    \cos(\theta/2) }
\pmatrix{\cosh\lambda & -\sinh\lambda  \cr
 -\sinh\lambda & \cosh\lambda}
\pmatrix{\cos(\theta/2) & \sin(\theta/2) \cr-\sin(\theta/2) &
  \cos(\theta/2) } .
\end{equation}
This form is called the Bargmann decomposition~\cite{barg47,hk88}.
This expression can be compressed to
\begin{equation}\label{bargd11}
[ABCA] = \pmatrix{\cosh\lambda \cos\theta & -\sinh\lambda -
    \cosh\lambda\sin\theta \cr
-\sinh\lambda + \cosh\lambda\sin\theta & \cosh\lambda \cos\theta } ,
\end{equation}
with two independent parameters.  We shall call this expression the
``Bargmann matrix.''
 \par
If $\sin\theta = \tanh\lambda$, the $ABCA$ matrix takes the
triangular form given in Eq.(\ref{wigd55}).  Then this special case
of the Bargmann decomposition is called the ``Iwasawa'' decomposition.
This form is also known to correspond to gauge transformations of
massless particles~\cite{knp86,hk88}.
\par
We shall see in this report that both Wigner and Bargmann decompositions
play essential roles in optics and space-time symmetries.   The question
then is whether one transformation can be translated into the other.  If
$\cosh\lambda \cos\theta$ is smaller than one, the off-diagonal elements
have opposite signs.  The Bargmann matrix of Eq.(\ref{bargd11}) will
be translated into Eq.(\ref{wigd11}) with
\begin{equation}
\cos\phi = \cosh\lambda \cos\theta, \qquad
e^\eta = \sqrt{\frac{\sin\theta - \tanh\lambda}{\sin\theta + \tanh\lambda}}
\end{equation}
If the diagonal element is greater than one, the matrix will be translated
into Eq.(\ref{wigd22}) with
\begin{equation}
\cosh\chi = \cosh\lambda \cos\theta, \qquad
e^\eta = \sqrt{\frac{\tanh\lambda - \sin\theta}{\tanh\lambda + \sin\theta}}
\end{equation}
These variable transformations are simple enough, but the real issue is
what happens when the diagonal element makes a transition from less-than-one
to greater-than-one.  If it becomes one, the lower left element of the
Bargmann matrix becomes zero, and the matrix becomes triangular.

%----------------------------------------------------------------------
\begin{figure}%[thb]
\centerline{\includegraphics[scale=0.6]{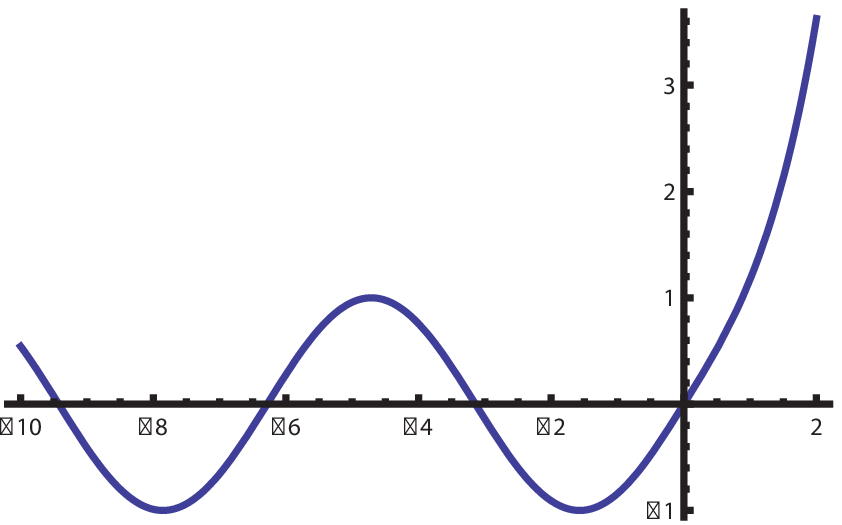}
\hspace{10mm}
\includegraphics[scale=0.6]{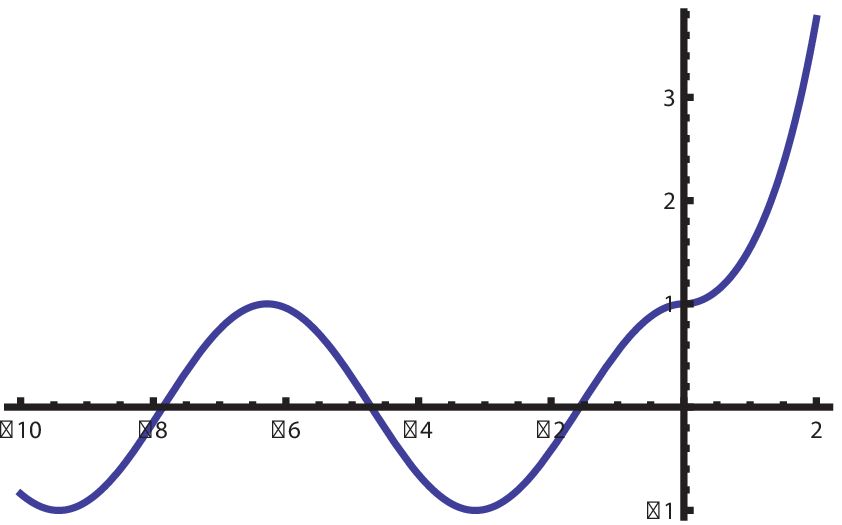}}
\caption{Transitions from $\sin$ to $\sinh$, and from $\cos$
to $\cosh.$  They are continuous transitions.  Their first derivatives
are also continuous, but the second derivatives are not.  Thus, they
are not analytic continuations.}\label{sincos}
\end{figure}
%----------------------------------------------------------------------

\par
In order to study this transition, we introduce a small number
\begin{equation}
\epsilon = \cosh\lambda \sin\theta - \sinh\lambda,
\end{equation}
and see what happens when it changes its sign.  When $\epsilon$ is
positive, the Bargmann matrix can be written as
\begin{equation}
\pmatrix{1 - \epsilon \sinh\eta\cosh\eta  &  -2\sinh\eta \cr
 \epsilon \cosh\eta &   1 - \epsilon \sinh\eta\cosh\eta} .
\end{equation}
If we let
\begin{equation}
\alpha = \sqrt{2\epsilon \sinh\eta \cosh\eta} , \qquad
   \beta = \sqrt{\frac{\epsilon \cosh\eta}{2\sinh\eta}},
\end{equation}
the matrix becomes
\begin{equation}
\pmatrix{1 - \alpha^2/2  & -\alpha/\beta \cr
        \alpha\beta & 1 - \alpha^2/2 } ,
\end{equation}
which can be decomposed into
\begin{equation}
 \pmatrix{1/\sqrt{\beta} & 0 \cr 0 & \sqrt{\beta}}
  \pmatrix{1 - \alpha^2/2  & -\alpha \cr
    \alpha & 1 - \alpha^2/2 }
 \pmatrix{ \sqrt{\beta} & 0 \cr 0 & 1/\sqrt{\beta}} .
\end{equation}
 \par
If $\epsilon$ is negative, we should define $\alpha$ and $\beta$
as
\begin{equation}
\alpha = \sqrt{-2\epsilon \sinh\eta \cosh\eta} , \qquad
   \beta = \sqrt{\frac{-\epsilon \cosh\eta}{2\sinh\eta}},
\end{equation}
and the matrix becomes
\begin{equation}
\pmatrix{1 + \alpha^2/2  & -\alpha\beta \cr
        - \alpha/\beta & 1 + \alpha^2/2 } ,
\end{equation}
which can be decomposed into
\begin{equation}
 \pmatrix{1/\sqrt{\beta} & 0 \cr 0 & \sqrt{\beta}}
  \pmatrix{1 + \alpha^2/2  & -\alpha \cr
    -\alpha & 1 + \alpha^2/2 }
 \pmatrix{ \sqrt{\beta} & 0 \cr 0 & 1/\sqrt{\beta}} .
\end{equation}
\par
It is clear now that the transition from the rotation-like Wigner
(with $\cos\phi$ as the diagonal element) to the squeeze-like matrix
(with $\cosh\chi$ as the diagonal element) is like the transitions
from $\sin\alpha$ to $\sinh\alpha$, and from $\cos\alpha$ to
$\cosh\alpha,$ as shown in Fig.~\ref{sincos}.  This is a continuous
transition, but not analytic.  The second derivative is not
continuous.

\section{Periodic System in Optics}\label{optsys}
Let us consider an optical beam going through multiple layers consisting
of two different refractive indexes.  This problem has been extensively
discussed in the literature.  The two-by-two matrix formulation of this
problem is given in the Appendix.
\par
When the beam goes through the first medium, we can use the matrix
\begin{equation}
    R\left(2\phi_1\right) = \pmatrix{\cos\phi_1 & -\sin\phi_1 \cr
                      \sin\phi_1 & \cos\phi_1} .
\end{equation}
For the second medium, we use $\phi_2$ instead of $\phi_1$.
\par
If the beam in the first medium hits the second medium, it is
partially transmitted and partially reflected.  According to
the Appendix, the boundary matrix takes the form
\begin{equation}
 B(\eta) = \pmatrix{e^{\eta/2} & 0 \cr 0 & e^{-\eta/2}} .
\end{equation}
This form is also given in Eq.(\ref{wigd66}) in connection with
the Wigner decomposition.  When the beam hits the first medium
from the second medium, the boundary matrix is $B(-\eta)$.
\par
\begin{figure}%[thb]
\centerline{\includegraphics[scale=0.4]{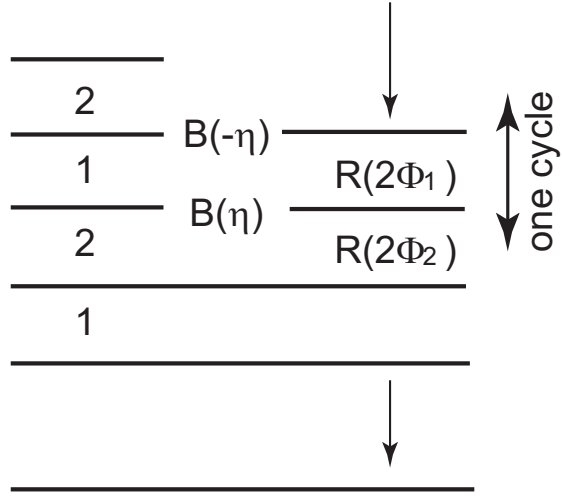}}
\caption{Optical layers.  There are phase-shift matrices for their
respective layers.  There is a boundary matrix for the transition from
the first to second medium, and its inverse applies to the transition
from the second to first medium.  The cycle starts from the middle of
the second layer.} \label{layer55}
\end{figure}

Let us consider the cycle which starts from the half-way in the
second medium and ends at the second medium as illustrated in
Fig~\ref{layer55}.  Then the beam transfer matrix becomes
\begin{equation}
[ABCA]  = R\left(\phi_2\right)\left[B(\eta)R\left(2\phi_1\right)
    B(-\eta)\right]
R\left(\phi_2\right) .
\end{equation}
The quantity inside the square bracket is takes a form the Wigner
decomposition, but it is sandwiched between two rotation matrix.
Our problem is to write the entire matrix chain as a Wigner
decomposition, convenient for the periodic system.  For this purpose,
we write the Wigner decomposition in the middle as a Bargmann decomposition
\begin{equation}
R(\theta) S(-2\lambda) R(\theta) ,
\end{equation}
where $\theta$ and $\lambda$ are determined by
\begin{eqnarray}
&{}& \cosh\lambda =
 (\cosh\eta)\sqrt{1 - \cos^{2}\phi_1\tanh^{2}\eta}, \nonumber
\\[2ex]
&{}& \cos\theta = {\cos\phi_1 \over (\cosh\eta)
      \sqrt{1 - \cos^{2}\phi_1\tanh^{2}\eta} } .
\end{eqnarray}

%------------------------------------------------------------------------

%------------------------------------------------------------------------

\par

We can now write $[ABCA]$ as
\begin{equation}
  R\left(\phi_2\right) R(\theta) S(-2\lambda)
  R(\theta) R\left(\phi_2\right) .
\end{equation}
Since $R(\theta) R\left(\phi_2\right) = R\left(\theta + \phi_2/2\right),$
the $[ABCA]$ matrix becomes a Bargmann decomposition of the form
\begin{equation}
[ABCA] = R(\theta^*) S(-2\lambda) R(\theta^*) ,
\end{equation}
with
\begin{equation}
\theta^{*} = \theta + \frac{1}{2}\phi_2 .
\end{equation}
\par
After matrix multiplications, the $[ABCA]$ matrix can be written as
\begin{equation}\label{bgd101}
\pmatrix{\cosh\lambda \cos\theta^*  &
       -\sinh\lambda - \cosh\lambda \sin\theta^* \cr
 -\sinh\lambda + \cosh\lambda \sin\theta^*  &
 \cosh\lambda \cos\theta^*}
\end{equation}

\par
For the multilayer system, we need a repeated application of this
from.  For this purpose, we have to convert this into the
Wigner decomposition:
\begin{equation}
   B(\eta^*) W(\tau^*) B(-\eta^*) .
\end{equation}
When $\cosh\lambda \cos\theta^*$ is smaller than one,
\begin{equation}
 W(\tau^*) = \pmatrix{\cos\phi^* & -\sin\phi^* \cr \sin\phi^* & \cos\phi^*},
\end{equation}
with
\begin{equation}
\cos\phi^* = \cosh\lambda \cos\theta^*, \qquad
\exp{(\eta^*)} = \sqrt{\frac{\sin\theta^* - \tanh\lambda}
                   {\sin\theta^* + \tanh\lambda}}
\end{equation}
If the diagonal element is greater than one, the $W(\tau^*)$ matrix is of the form
\begin{equation}
 W(\tau*) = \pmatrix{\cosh\chi^* & -\sinh\chi^* \cr -\sinh\chi^* & \cosh\chi^*},
\end{equation}
with
\begin{equation}
\cosh\chi^* = \cosh\lambda \cos\theta^*, \qquad
\exp{(\eta)} = \sqrt{\frac{\tanh\lambda - \sin\theta^*}
        {\tanh\lambda + \sinh\theta^*}}
\end{equation}
The situation is similar if the diagonal element is one, and the matrix
is triangular.  We can use this Wigner decomposition to calculate
\begin{equation}
  [B(\eta^*) W^(\tau*) B(-\eta^*)]^N = B(\eta^*)
             \left[W^(\tau^*)\right]^N B(-\eta)
\end{equation}
\par
It is seen that this periodic system contains the Wigner, Bargmann, and
Iwasawa decompositions in its natural language, and it forces us to bring
the $[ABCA]$ matrix into its Wigner decomposition for its repeated
applications in the periodic system.   In Sec~\ref{little}, we shall
study what this Wigner decomposition means in symmetries in particle
physics.

\section{Space-time Symmetries}\label{little}
In 1939~\cite{wig39}, Eugene Wigner considered subgroups of the Lorentz
group whose transformations leave the given momentum of a particle
invariant.  These subgroups are called Wigner's little groups.  While
leaving the momentum invariant, the little group transforms internal
space-time variables.  For instance, if the particle is at rest, rotations
do not change its momentum, but they can change the orientations of the
particle spin.  In this section, we formulate Wigner's little group
using the technique of decompositions discussed in Sec.~\ref{decom} and
Sec.~\ref{optsys}.
\par

In Sec.~\ref{decom}, we studied the decompositions of the $[ABCD]$
into single-parameter matrices, using the two-by-two matrices
$B(\eta), R(\phi),$ and $S(\chi)$.
It is known that these two-by-two matrices correspond to the
four-by-four Lorentz-transformation matrices applicable to the
Minkowskian space-time coordinate variables $(z, y, z, t)$~\cite{knp86}.
The two-by-two matrix $B(\eta)$ can also be written as the
the four-by-four matrix for the Lorents boost along the $z$ direction:
\begin{equation}\label{mat53}
B(\eta) = \pmatrix{e^{\eta/2} & 0 \cr 0 & e^{-\eta/2}}  \rightarrow
  \pmatrix{1 & 0 & 0 & 0 \cr
   0 & 1 & 0 & 0   \cr
   0 & 0 & \cosh\eta & \sinh\eta \cr
        0 & 0 & \sinh\eta & \cosh\eta} ,
\end{equation}
which performs a Lorentz boost along the $z$ direction.  The
two-by-two matrix $R(\phi)$ an be translated into
\begin{equation}\label{rot55}
 R(\phi) = \pmatrix{\cos(\phi/2) & -\sin(\phi/2) \cr \sin(\phi/2)
  & \cos(\phi/2)} \rightarrow
  \pmatrix{\cos\phi & 0 & \sin\phi & 0
   \cr 0 & 1 & 0 & 0 \cr
    -\sin\phi & 0 & \cos\phi & 0 \cr 0 & 0 & 0 & 1} ,
\end{equation}
which performs a rotation around the $y$ axis.  The $S(\chi)$
corresponds to
\begin{equation}
S(\chi) = \pmatrix{\cosh(\chi/2) & \sinh(\chi/2) \cr
  \sinh(\chi/2) & \cosh(\chi/2)} \rightarrow
  \pmatrix{\cosh\chi & 0 & 0 & \sinh\chi  \cr
   0 & 1 & 0 & 0   \cr   0 & 0 & 1 & 0 \cr
   \sinh\chi & 0 & 0 & \cosh\chi } ,
\end{equation}
which performs a Lorentz boost along the $x$ direction.
\par
It is possible now to translate the contents of Secs.~\ref{decom}
and~\ref{optsys} into the language of four-by-four Lorentz
transformation matrices applicable to the Minkowski space of
$(z,y,z,t)$.  In this convention, the momentum-energy four-vector
is $\left(p_x, p_y, p_z, E\right).$  If the particle moves along
the $z$ direction, this four-vector
becomes
\begin{equation}\label{mom11}
\left(0, 0, p, \sqrt{p^2 + m^2}\right) ,
\end{equation}
in the unit system where $c = 1,$ where $m$ is the particle mass.  We
can obtain this four-vector by boosting a particle at rest with the
four-momentum
\begin{equation}\label{mom22}
(0, 0, 0, m) ,
\end{equation}
using the four-by-four boost matrix given in Eq.(\ref{mat53}), with
\begin{equation}
\tanh(\eta) = \frac{p}{\sqrt{p^2 + m^2}} .
\end{equation}

\par

\begin{figure}%[thb]
\centerline{\includegraphics[scale=0.45]{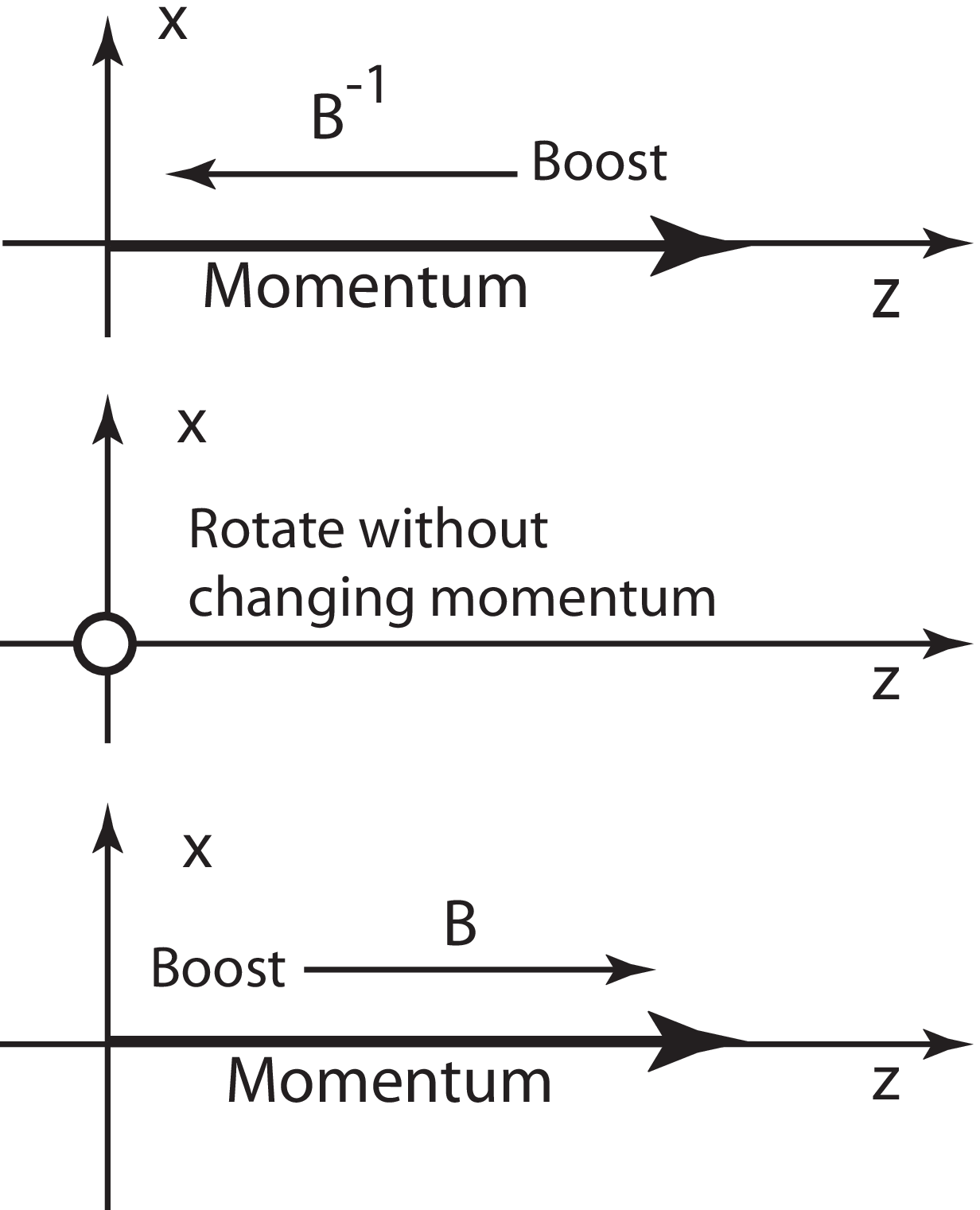}
\hspace{10mm}
\includegraphics[scale=0.34]{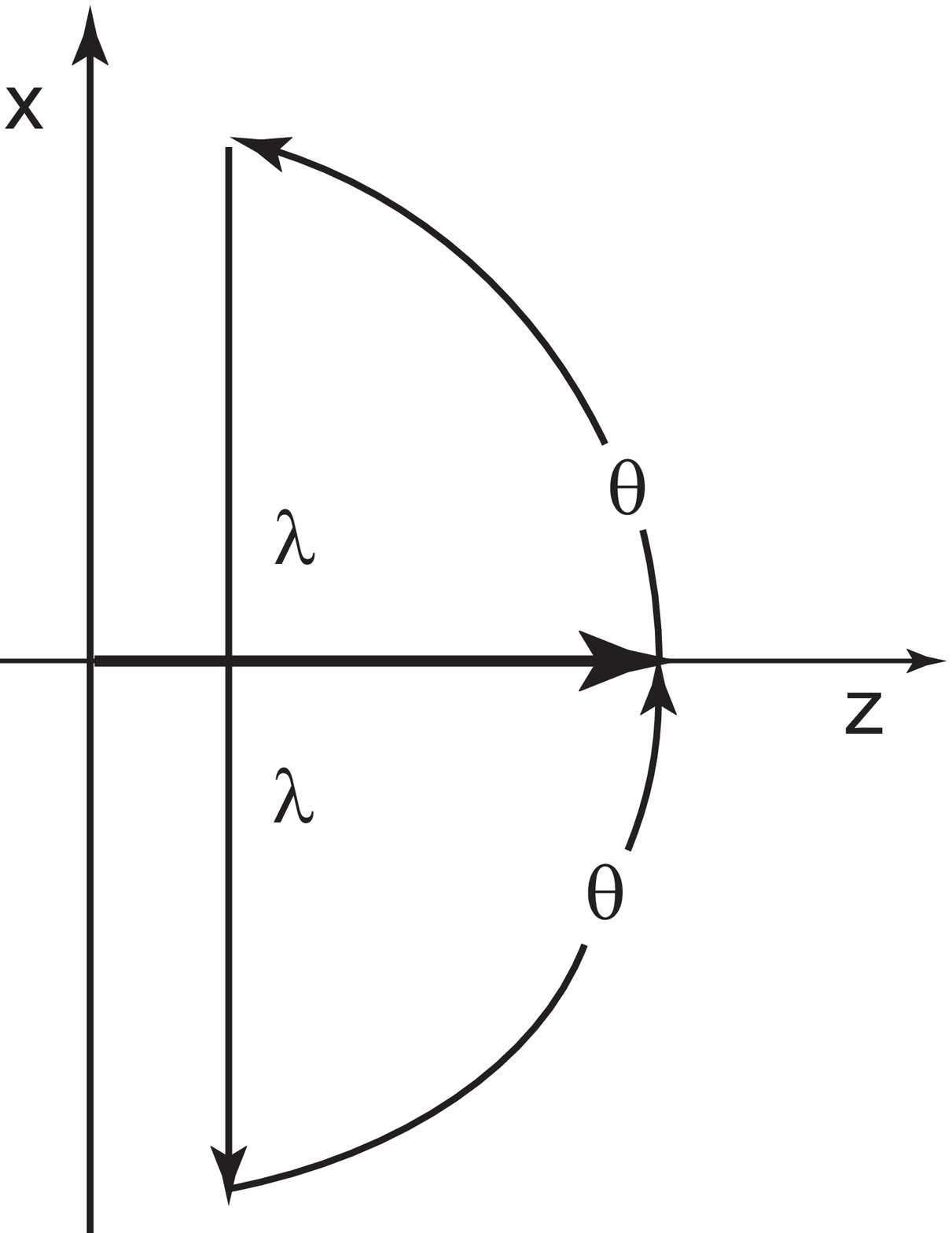}}
\caption{Illustrations of the Wigner decomposition (left) and the
  Bargmann decomposition.   In both cases, the net transformation
  leaves the four-momentum of the particle unchanged.  }\label{bw22}
\end{figure}

Now the four-momentum of Eq.(\ref{mom22}) is invariant under the
rotation matrix
\begin{equation}\label{rot56}
 R(2\phi) = \pmatrix{\cos(2\phi z) & 0 & \sin(2\phi) & 0
   \cr 0 & 1 & 0 & 0 \cr
    -\sin(2\phi) & 0 & \cos(2\phi) & 0 \cr 0 & 0 & 0 & 1} .
\end{equation}
Thus, the matrix
\begin{equation}\label{wlg11}
B(\eta) R(2\phi) B(-\eta)
\end{equation}
leaves the four-momentum of Eq.(\ref{mom11}) invariant.  After
making this rotation, we can bring the momentum to its initial
state by boosting it by $B(\eta)$.  The net effect is the
momentum-preserving transformation.  This set of transformations
is illustrated in Fig.~\ref{bw22}, and corresponds to the
Wigner decomposition.

\par
If the particle has a space-like momentum, we can start with the
four-momentum
\begin{equation}
  (0, 0, p, E) ,
\end{equation}
where $E$ is smaller than $p$, which it can be brought to the
Lorentz frame where the four-vector becomes
\begin{equation}\label{mom33}
  (0, 0, p, 0) .
\end{equation}
The boost matrix takes form of Eq.(\ref{mat53}), with
\begin{equation}
\tanh(\eta) = \frac{E}{p} .
\end{equation}
The four-momentum of Eq.(\ref{mom33}) is invariant under the boost
\begin{equation}
S(-2\chi) = \pmatrix{\cosh(2\chi) & 0 & 0 & -\sinh(2\chi)  \cr
   0 & 1 & 0 & 0   \cr   0 & 0 & 1 & 0 \cr
   -\sinh(2\chi) & 0 & 0 & \cosh(2\chi) }
\end{equation}
along the $x$ direction.

%----------------------------------------------------------------------
%----------------------------------------------------------------------

\par
Let us finally consider a massless particle with it four-momentum
\begin{equation}
(0, 0, p, p) .
\end{equation}
It is invariant under the rotation around the $z$ axis.  In addition, it
is invariant under the transformation
\begin{equation}\
\pmatrix{1 & 0 & -2\gamma & 2\gamma \cr 0 & 1 & 0 & 0 \cr
               2\gamma & 0 &  1 -2\gamma^2 & 2\gamma^2 \cr
               2\gamma & 0 &   -2\gamma^2 & 1 + 2\gamma^2 }.
\end{equation}
This four-by-four matrix has a stormy history~\cite{knp86,kiwi90jm},
but the bottom line is that it corresponds to the triangular matrix
of Eq.(\ref{wigd55}), and the variable $\gamma$ performs gauge
transformations.
\par
We can obtain this massless case from the massive or imaginary case
using the limiting procedure spelled out in Sec.~\ref{decom}.  This
procedure is widely known as the group contraction in the literature.
\par
As for the Bargmann decomposition, let us go to Fig.~\ref{bw22}.
The particle moving along the $z$ direction can be rotated first
by $R(\theta)$.  It can then be  boosted along the negative $x$
axis, and then rotated again by $R(\theta)$ to the original position.
Indeed, this is a momentum-preserving transformation.  The net
transformation can be written as
\begin{equation}
 R(\theta) S(-2\lambda) R(\theta) .
\end{equation}
This Bargmann decomposition is applicable to all three cases of
the momentum~\cite{hk88}.

\par
In addition, Wigner's little group allows rotations around the
momentum which does not change it.  This extra degree of freedom
does not affect the description of the symmetries given in this
section~\cite{hks86jm}

\section*{Concluding Remarks}
In this report, we started with a two-by-two matrix with real
elements, but we were led to consider three classes of
equi-diagonal matrices, with their traces less than two, equal
to two, and greater than two.  From the mathematical point of
view, this process is the construction of group representations
according to conjugate classes.
\par
These conjugate classes correspond to Wigner's little groups for
the internal space-time symmetries for massive, massless, and
imaginary-mass particles in the Lorentz-covariant world.  Indeed,
this aspect is a very ``unreasonable agreement'' between
mathematics and physics.

\par
It was noted that these equi-diagonal matrices as products of
three one-parameter matrices, resulting in the Wigner, Bargmann,
and Iwasawa decompositions.  It is noted further that the optical
periodic systems, such as multilayer optics, can perform these
decompositions.  Thus, the optical periodic system speaks the
language of the fundamental symmetries for elementary particles
in Einstein's Lorentz-covariant world.

\section*{Appendix}
The $N$-layer optics starts with the boundary matrix of the
form~\cite{monzon00,gk01}
\begin{equation}\label{bd11}
B(\eta) = \pmatrix{\cosh(\eta/2) & \sinh(\eta/2) \cr \sinh(\eta/2) &
\cosh(\eta/2) } ,
\end{equation}
which, as illustrated in Fig.~\ref{layer55}, describes the transition
from $medium~2$ to $medium~1$, taking into account both the
transmission and reflection of the beam.  As the beam goes through
the $medium~1$, the beam undergoes the phase shift represented by
the matrix
\begin{equation}\label{ps11}
P(\phi_1) = \pmatrix{e^{-i\phi_1} & 0 \cr 0 & e^{i\phi_1}} .
\end{equation}
When the wave hits the surface of the second medium, the corresponding
matrix is
\begin{equation}\label{bd22}
B(-\eta) = \pmatrix{\cosh(\eta/2) & -\sinh(\eta/2) \cr -\sinh(\eta/2) &
\cosh(\eta/2) } ,
\end{equation}
which is the inverse of the matrix given in Eq.(\ref{bd11}).
Within the second medium, we write the phase-shift matrix as
\begin{equation}\label{ps22}
P(\phi_2) = \pmatrix{e^{-i\phi_2} & 0 \cr 0 & e^{i\phi_2}} .
\end{equation}
Then, when the wave goes through one cycle starting from the
midpoint in the second medium, the beam transfer matrix becomes
\begin{eqnarray}\label{m1}
&{}&
M_1 = \pmatrix{e^{-i\phi_2/2} & 0 \cr 0 & e^{i\phi_2/2}}
\pmatrix{\cosh(\eta/2) &
\sinh(\eta/2) \cr \sinh(\eta/2) & \cosh(\eta/2) }
\pmatrix{e^{-i\phi_1} & 0 \cr 0 & e^{i\phi_1}} \nonumber \\[2ex]
&{}& \hspace{10mm} \times
\pmatrix{\cosh(\eta/2) & -\sinh(\eta/2) \cr
         -\sinh(\eta/2) & \cosh(\eta/2) }
\pmatrix{e^{-i\phi_2/2} & 0 \cr 0 & e^{i\phi_2/2}} .
\end{eqnarray}
This arrangement of the matrices is illustrated in Fig.~\ref{layer55}

\par
The $M_{1}$ matrix Eq.(\ref{m1}) contains complex numbers, but we
are interested in carrying out calculations with real matrices.
This can be done by means of a conjugate or similarity
transformation~\cite{gk01}.  Let us next  consider the matrix
\begin{equation}
C = {1 \over 2} \pmatrix{1 & 1 \cr -1 & 1} \pmatrix{1 & i \cr i & 1}
 = {1 \over \sqrt{2}} \pmatrix{e^{i\pi/4} &  e^{i\pi/4} \cr
-e^{-i\pi/4} & e^{-i\pi/4}} .
\end{equation}
This matrix and its inverse can be written as
\begin{equation}
C = { e^{i\pi/4} \over \sqrt{2}} \pmatrix{ 1 & 1 \cr i  & -i},
\qquad
C = { e^{-i\pi/4} \over \sqrt{2}} \pmatrix{ 1 & -i \cr 1  & i} .
\end{equation}
\par
We can then consider the conjugate transform of the $M_1$ matrix
\begin{equation}\label{conju11}
M_{2} = C~M_{1}~C^{-1} ,
\end{equation}
with
\begin{eqnarray}\label{core22}
&{}& M_{2} = \pmatrix{\cos(\phi_2/2) & -\sin(\phi_2/2) \cr
\sin(\phi_2/2) &  \cos(\phi_2/2) }
\pmatrix{e^{\eta/2} & 0 \cr 0 & e^{-\eta/2} }
\pmatrix{\cos\phi_1 & -\sin\phi_1 \cr
        \sin\phi_1 &  \cos\phi_1} \nonumber \\[2ex]
&{}& \hspace{16mm} \times
\pmatrix{ e^{-\eta/2} & 0 \cr 0 & e^{\eta/2}}
\pmatrix{\cos(\phi_2/2) & -\sin(\phi_2/2) \cr
\sin(\phi_2/2) &  \cos(\phi_2/2) }  .
\end{eqnarray}
The conjugate transformation of Eq.(\ref{conju11}) changes
the boundary matrix $B(\eta)$ of Eq.(\ref{bd11}) to a
squeeze matrix
\begin{equation}\label{sq11}
B(\eta) = \pmatrix{ e^{\eta/2} & 0 \cr 0 & e^{-\eta/2} } ,
\end{equation}
and the phase-shift matrices $P(\phi_1)$ of Eq.(\ref{ps11}) and
Eq.(\ref{ps22}) to rotation matrices
\begin{equation}\label{rot22}
R(2\phi_{i}) = \pmatrix{\cos(\phi_{i}) & -\sin(\phi_{i}) \cr
\sin(\phi_{i}) &  \cos(\phi_{i}} ,
\end{equation}
with $i = 1,~2$.


\begin{thebibliography}{99}

\bibitem{theo79}
P. S. Theocaris and E. E. Gdoutos, {\em Matrix Theory of
Photoelasticity}(Springer-Verlag, Berlin, 1979).


\bibitem{wig39}
E. Wigner,  Ann. Math. {\bf 40}, 149 (1939).

\bibitem{knp86}
Y. S. Kim and M. E. Noz, {\em Theory and Applications of the
Poincar\'e Group} (Reidel, Dordrecht, 1986).

\bibitem{kiwi90jm}
Y. S. Kim and E. P. Wigner, J. Math. Phys. {\bf 31}, 55 (1990).

\bibitem{hkn99ajp}
D. Han, Y. S. Kim, and M. E. Noz, Am. J. Phys. {\bf 67} 61 (1999).

\bibitem{yuen76}
H. P. Yuen, Phys. Rev. A {\bf 13}, 2226 (1976).

\bibitem{dir49}
P. A. M. Dirac, Rev. Mod. Phys. {\bf 21}, 392 (1949).

\bibitem{dir63}

P. A. M. Dirac, J. Math. Phys. {\bf 4}, 901 (1963).


\bibitem{barg47}
V. Bargmann,  Ann. Math. {\bf 48}, 568--640 (1947).


\bibitem{gk01}
E. Georgieva and Y. S. Kim  Phys. Rev. E {\bf 64} 026602 (2001).


\bibitem{hamer62}
M. Hamermesh, {\em Group Theory and Its Application to Physical
Problems} (Addison-Wesley, Reading Massachusetts, U.S.A.).

\bibitem{knp91}
Y. S. Kim and M. E. Noz, {\em Phase Space Picture of Quantum
Mechanics} (World Scientific, Singapore, 1991).

\bibitem{bk10jmo}
S. Baskal and Y. S. Kim, J. Mod. Opt. {\bf 57}, 1251 (2010).

\bibitem{hk88}
D. Han and Y. S. Kim, Phys. Rev. A {\bf 37}, 4494 (1988).

\bibitem{bk09}
S. Baskal and Y. S. Kim, J. Opt. Soc. Am. A {\bf 26}, 3049 (2009)

\bibitem{hks86jm}
D. Han, Y. S. Kim, and D. Son, J. Math. Phys. {\bf 27}, 2228 (1986).


\bibitem{monzon00}
J. J. Monz\'on and L. L. S\'anchez-Soto, J. Opt. Soc. Am. A, {\bf 17},
1475 (2000).

\end{thebibliography}
\end{document}